\documentclass[11pt]{article}
\usepackage{fancyhdr,psfig,lastpage}
\usepackage{amsmath}
\usepackage{latexsym,cite}
\begin{document}

\vspace*{0.8cm}
\begin{center}
{\Large \bf Directed motion of Brownian particles \\ with internal
  energy depot 
}\\[10mm]

{\large Benno Tilch$^{1}$, Frank Schweitzer$^{2\star}$, Werner Ebeling$^{2}$}\\[4mm]

\begin{quote}
\begin{itemize}
\item[$^{1}$]{\it II. Institut f\"ur Theoretische Physik, Universt\"at
  Stuttgart, Pfaffenwaldring 57/III, D-70550 Stuttgart, Germany} 
\item[$^{2}$]{\it Institut f\"ur Physik, Humboldt--Universit\"at zu
  Berlin, Unter den Linden 6, D-10099 Berlin, Germany} 

\item[$\star$] Corresponding author: frank@physik.hu-berlin.de 
\end{itemize}
\end{quote}
\end{center}

\begin{abstract}
  A model of Brownian particles with the ability to take up energy from
  the environment, to store it in an internal depot, and to convert
  internal energy into kinetic energy of motion, is discussed. The
  general dynamics outlined in Sect. 2 is investigated for the
  deterministic and stochastic particle's motion in a non-fluctuating
  ratchet potential. First, we discuss the attractor structure of the
  ratchet system by means of computer simulations. Dependent on the
  energy supply, we find either periodic bound attractors corresponding
  to localized oscillations, or one/two unbound attractors corresponding
  to directed movement in the ratchet potential. Considering an ensemble
  of particles, we show that in the deterministic case two currents into
  different directions can occur, which however depend on a
  supercritical supply of energy. Considering stochastic influences, we
  find the current only in one direction. We further investigate how the
  current reversal depends on the strength of the stochastic force and
  the asymmetry of the potential. We find both a critical value of the
  noise intensity for the onset of the current and an optimal value where
  the net current reaches a maximum. Eventually, the dynamics of our
  model is compared with other ratchet models previously suggested.
\end{abstract}


\newcommand{\bbox}[1]{\mbox{\boldmath $#1$}}
\newcommand{\mean}[1]{\left\langle #1 \right\rangle}
\newcommand{\abs}[1]{\left| #1 \right|}
\newcommand{\la}{\langle}
\newcommand{\ra}{\rangle}
\newcommand{\RA}{\Rightarrow}
\newcommand{\tet}{\vartheta}
\newcommand{\eps}{\varepsilon}
\renewcommand{\phi}{\varphi}
\newcommand{\bib}[4]{\bibitem{#1} {\rm #2} (#4): #3.}
\newcommand{\ul}[1]{\underline{#1}}
\newcommand{\non}{\nonumber \\}
\newcommand{\no}{\nonumber}
\newcommand{\sect}[1]{Sect. \ref{#1}}
\newcommand{\eqn}[1]{Eq. (\ref{#1})}
\newcommand{\eqs}[2]{Eqs. (\ref{#1}), (\ref{#2})}
\newcommand{\pic}[1]{Fig. \ref{#1}}
\newcommand{\vol}[1]{{\bf #1}}
\newcommand{\et}{{\it et al.}}
\newcommand{\fn}[1]{\footnote{ #1}}
\newcommand{\name}[1]{{\rm #1}}
\newcommand{\D}{\displaystyle}
\newcommand{\T}{\textstyle}
\newcommand{\SC}{\scriptstyle}
\newcommand{\SSC}{\scriptscriptstyle}

\section{Introduction}

Among the phenomena presently investigated in cell biology
is the generation of directed movement of ``particles'' (e.g. kinesin or
myosin molecules) along periodic structures (e.g. microtubules or actin
filaments) \cite{finer-nat-94} in the \emph{absense of a
  macroscopic force}, which may have resulted from temperature or
concentration gradients. In order to reveal the microscopic mechanisms
resulting in directed movement, different physical ratchet models have
been proposed \cite{maddox-nat-94b}, such as {\em forced
  thermal ratchets} \cite{magnasco-93}, or {\em stochastic ratchets}
\cite{luczka-bart-haenggi-95,millonas-dykman-94}, or {\em fluctuating
  ratchets}
\cite{rousselet-nat-94,zuercher-doering-93}. These models are based on different types of long-range order
correlations, e.g. a periodic asymmetric potential, or a time-dependent
change of noise, or a fluctuating energy barrier (cf. also \sect{4}).
They have in common to transfer the undirected motion of Brownian
particles into a directed motion, hence, the term {\em Brownian
  rectifiers} \cite{haenggi-bart-96} has been established.

The ratchet models have recently proved their value for describing
Brownian machines or {\em molecular motors}
\cite{magnasco-94,astumian-bier-94,juelicher-prost-95}, which convert
chemical energy into mechanical motion. The model discussed in this
paper, aims to add a new perspective to this problem. Our model is based
on {\em active Brownian particles}
\cite{fs-lsg-94,steuern-et-94,lsg-mieth-rose-malch-95,
  fs-agent-97,fs-eb-tilch-98-let,eb-fs-tilch-98}, which are Brownian
particles with the ability to take up energy from the environment, to
store it in an internal depot and to convert internal energy to perform
different activities, such as metabolism, motion, change of the
environment, or signal-response behavior.  In this paper, we focus on the
energetic aspects of the motion of these active Brownian particles in a
specific potential and neglect the possible changes of the environment.

Different from ``usual'' Brownian particles which only move
\emph{passively} caused by the influence of a stochastic force, active
Brownian particles have the capability of \emph{active motion}. This
means for instance the possibility of accelerated motion, provided there
is a supercritical supply of energy from the environment.  In general, we
can think of different mechanisms to pump additional energy into the
motion of the particles. For example, a complex friction coefficient,
$\gamma(\bbox{r},\bbox{v})$, can be considered, which is a space- and /or
velocity dependent function, which can be also negative under certain
conditions.  Space-dependent negative friction has been discussed in
\cite{steuern-et-94}, while different ansatzes for a complex
velocity-dependent friction coefficient are investigated in
\cite{eb-fs-tilch-98,ebeling-et-99}. It can be shown that above a
critical influx of energy the Brownian particles are able to perform a
selfsustained limit cycle motion.

In our model, the Brownian particles are not just pumped by additional
energy, but also have the capability to store energy in an internal
depot. This means an additional degree of freedom for the particle, which
may allow a simplified description of active biological motion on the
micro-level \cite{eb-fs-tilch-98}. Considering for instance a spatially
inhomogeneous supply of energy, or the presence of obstacles, we have
shown that the motion of active Brownian particles in the two-dimensional
space can become rather complex \cite{fs-eb-tilch-98-let}. We have found
e.g. intermittent types of motion where cycles of accelerated motion
alternate with cycles of ``simple'' Brownian motion, or deterministic
chaos in the presence of reflecting obstacles.

In this paper, we investigate the motion of an ensemble of Brownian
particles with internal energy depot in a ratchet potential, i.e.  a
periodic potential which lacks the reflection symmetry.  In \sect{2}, we
discuss the basic features of our model 
and derive the equations of motion and the stationary solutions for the
case of a linear potential. 
In \sect{3}, we specify the external potential as a
non-fluctuating ratchet potential. By means of computer simulations, we
investigate the attractor structure of the particles motion and show the
existence of bound and unbound attractors dependent on the energy supply.
Further, we 
compare the deterministic motion of the particle ensemble with the
stochastic motion. We show that in the deterministic case a positive net
current occurs, which changes its direction in the presence of stochastic
influences. In order to investigate the current reversal in more detail,
in \sect{4} the dependence of the net current on the energy supply and
the asymmetry of the potential is calculated both for the deterministic
and the stochastic case. In \sect{5}, we compare our model with different
ratchet models suggested previously, and point to some differences. 

\section{Model of Pumped Brownian Dynamics}  
\label{2}
\subsection{Equations of Motion and Distribution Function}
\label{2.1}

The model of active Brownian motion investigated in this paper is based
on the idea \cite{fs-eb-tilch-98-let,eb-fs-tilch-98} that the particles
have the capability to store energy in an internal energy depot, $e(t)$,
which may be changed due to three different processes:
\begin{enumerate}
\item[(i)] gain of energy from the environment, where $q(\bbox{r})$ is the
  flux of energy into the depot. If the availability of energy is
  inhomogeneously distributed, the energy flux may depend on the space
  coordinate, $\bbox{r}$.
\item[(ii)] loss of energy due to internal dissipation, which is
  assumed to be proportional to the internal energy. Here the rate of
  energy loss, $c$, is assumed to be constant.
\item[(iii)] conversion of internal energy into kinetic energy with a
  rate $d(\bbox{v})$, which should be a function of the actual velocity,
  $\bbox{v}$, of the particle.
\end{enumerate}
The resulting balance equation for the energy depot is given by:
\begin{equation}
\frac{d}{dt} e(t) = q(\bbox{r}) - c\;e(t) - d(\bbox{v})\;e(t)
\label{dep0}
\end{equation}
A simple ansatz for $d(\bbox{v})$ reads: 
\begin{equation}
d(\bbox{v}) = d_2 v^2\;\;; \quad d_2>0
\label{dv}
\end{equation}
If we consider the motion of a particle with mass $m$ in an
\emph{external potential}, $U(\bbox{r})$, the balance equation for the
mechanical energy $E_{0}$ of the particle reads:
\begin{equation}
\frac{d}{dt} E_{0}(t) = \frac{d}{dt}\left(\frac{1}{2}m\bbox{v}^2 +
  U(\bbox{r})\right) = \Big(\eta(v^{2})\,d_2\, e(t) - \gamma_{0}\Big)\, \bbox{v}^2
\label{mechen}
\end{equation}
It means that $E_0$ can be changed by two processes:
\begin{enumerate}
\item increase of the kinetic energy by conversion of depot energy, where
  $\eta(v^{2})\leq 1$ is the \emph{conversion efficiency}
\item decrease by disspation of energy due ot the friction, $\gamma_{0}$,
  of the moving particle.
\end{enumerate}
Based on \eqn{mechen}, we
postulate a stochastic equation of motion for pumped Brownian
particles \cite{fs-eb-tilch-98-let,eb-fs-tilch-98} which is consistent
with the known \name{Langevin} equation of Brownian motion:
\begin{equation} 
m\bbox{\dot{v}} + \gamma_0 \,\bbox{v} 
+ \bbox{\nabla} U(\bbox{r}) = \eta(v^{2})d_2 e(t) \bbox{v} + 
\sqrt{2 D}\;\bbox{\xi}(t) 
\label{langev-dep}
\end{equation}
The right-hand side of the \name{Langevin} equation contains \emph{two
  driving forces} for the motion:
\begin{enumerate}
\item the acceleration in the direction of movement,
  $\bbox{e}_v=\bbox{v}/v$, due to the conversion of internal into kinetic
  energy,
\item a stochastic force with the strength $D$ and white-noise
  fluctuations: $\langle\xi(t)\xi(t')\rangle = \delta(t-t')$.
\end{enumerate}
According to the fluctuation-dissipation theorem, we assume that $D$ can be
expressed as:
\begin{equation}
  \label{fluct-diss}
D=k_{B}T\gamma_{0}  
\end{equation}
Because the loss of energy resulting from friction and the gain of energy
resulting from the stochastic force are compensated in the average, the
balance equation for the mechanical energy, \eqn{mechen}, reads for the
\emph{stochastic case}:
\begin{equation}
\mean{
\frac{d}{dt} \left(\frac{1}{2}m\dot{\bbox{r}}^2 + U(\bbox{r})\right)} 
= \mean{\eta(v^{2}) d_2 e(t)\, \dot{\bbox{r}}^2}
\label{e0dt-2}
\end{equation}
The presence of an internal energy depot means an additional degree of
freedom for the particle, which extends the phase space
$\Gamma=\{\bbox{r},\bbox{v},e\}$. Let $p(\bbox{r},\bbox{v},e,t)$ denote
the probability density to find the particle at time $t$ at location
$\bbox{r}$ with velocity $\bbox{v}$ and internal depot energy $e$. The
time-dependent change of the probability density can be described by a
\name{Fokker-Planck} equation, which has to consider now both the
\name{Langevin} \eqn{langev-dep} for the motion of the particle, and the
equation for the energy depot, \eqn{dep0}:
\begin{eqnarray}
\label{fpe-d2e}
\frac{\partial p(\bbox{r},\bbox{v},e,t)}{\partial t} & =&
\frac{\partial}{\partial \bbox{v}} \left\{
\frac{\gamma_{0}-\eta\, d_{2}e}{m}\,\bbox{v}\,
p(\bbox{r},\bbox{v},e,t) + \frac{D}{m^{2}} \, 
\frac{\partial p(\bbox{r},\bbox{v},e,t)}{\partial \bbox{v}}\right\} \non &&
 - \bbox{v}\, \frac{\partial p(\bbox{r},\bbox{v},e,t)}{\partial \bbox{r}}
+ \frac{1}{m}\,\bbox{\nabla} U(\bbox{r}) \,\frac{\partial
p(\bbox{r},\bbox{v},e,t)}{\partial\bbox{v}} \non &&
- \frac{\partial}{\partial e} \Big[
q(\bbox{r}) -c\, e -\eta\, d_{2}\bbox{v}^{2}e\Big]\,p(\bbox{r},\bbox{v},e,t)
\end{eqnarray}
For this probability density, the following normalization
condition holds:
\begin{equation}
  \label{norm-dep}
  \int_{-\infty}^{\infty} dr\; \int_{-\infty}^{\infty} dv
 \int_{0}^{\infty} de \; p(\bbox{r},\bbox{v},e,t) = 1
\end{equation}
In the following section, we will give the stationary solution for
$p(\bbox{r},\bbox{v},e,t)$ for a particular potential. A more detailed
discussion of the features of the probability density for pumped Brownian
particles is given in \cite{ebeling-et-99}.

\subsection{Motion in a linear Potential}
\label{2.2}
In the following, we restrict the discussion to the \emph{one-dimensional
  case}, i.e. the space coordinate is given by $x$. Further the mass of
the particle is set to $m=1$ and the flux of energy into the internal depot
of the particle is assumed as constant: $q(\bbox{r})=q_0$.
For the potential, we assume a linear function, hence the resulting
force is a constant:
\begin{equation}
    \label{u-x-f}
U(x)=ax \;;\quad F=-\nabla\,U = -a =const.   
\end{equation}
Then, the dynamics for the pumped Brownian motion is described by the
following set of equations:
\begin{eqnarray}
\label{model}
\dot{x} & = & v \non 
\dot{v} & = & - \Big(\gamma_{0} -\eta(v^{2}) d_2 e(t)\Big)v 
+ F + \sqrt{2D} \xi(t) \\ 
\mu\,\dot{e} & = & q_0 - c e - d_2 v^2 e \no
\end{eqnarray}
Here, we have introduced a formal parameter $\mu$ which can be used to
describe the time scale of relaxation of the internal energy depot.  The
limit \mbox{$\mu \rightarrow 0$} describes a very fast adaptation of the
depot.  Assuming further an ideal efficiency, $\eta=1$,
we get as an adiabatic approximation for the energy depot:
\begin{equation}
  \label{e0}
  e_{0}=\frac{q_{0}}{c+d_{2}v^{2}}
\end{equation}
This allows us to rewrite the equations of motion as: 
\begin{equation}
\label{model-2}
\dot{x} =  v \;;\quad
\dot{v}  =  - \gamma(v^{2})\,v
+ F + \sqrt{2D} \xi(t) 
\end{equation}
where $\gamma(v^{2})$ is a non-linear friction function:
\begin{equation}
  \label{gamma-v2}
 \gamma(v^{2})=\gamma_{0}- \frac{d_{2}\,q_{0}}{c+d_{2}v^{2}} 
\end{equation}
In the limit of large velocities, $\gamma(v^{2})$ approaches the normal
friction coefficient, $\gamma_{0}$, whereas in the limit of small
velocities a negative friction occurs, as an additional source of energy
for the Brownian particle. Hence, slow particles are accelerated, while
the motion of fast particles is damped. A more detailed discussion of
$\gamma(v^{2})$ is given in \cite{ebeling-et-99}. 

A similar discussion holds if we assume that the value of the energy
depot $e_{0}$ is constant and the conversion efficiency $\eta(v^{2})$
decreases as follows with increasing velocity:
\begin{equation}
  \label{eta-v2}
  \eta(v^{2})=\frac{\eta_{1}}{1+\eta_{2}v^{2}}
\end{equation}
where $\eta_{1}$ and $\eta_{2}$ are constants. Then we find for the
non-linear friction function with $\eta_{0}=\eta_{1}d_{2}$: 
\begin{equation}
  \label{gamma-v2-eta}
 \gamma(v^{2})=\gamma_{0}- \frac{\eta_{0}\,e_{0}}{1+\eta_{2}v^{2}} 
\end{equation}
which is a special case of \eqn{gamma-v2}.  In the following, we will
assume a constant conversion efficiency, $\eta(v^{2})=1$, and pay
particular attention to the conversion parameter $d_{2}$, instead.

If we first discuss the \emph{overdamped limit}, we can assume a
fast relaxation of the velocities, in which case
the set of equations, (\ref{model}), can be further reduced to:
\begin{equation}
\label{x-overd}
v(t) = - \frac{1}{\gamma_{0} 
- d_2 e_{0}}\;F 
+ \frac{\sqrt{2k_{B}T\gamma_{0}}}{\gamma_{0}-d_2e_{0}} \; \xi(t)
\end{equation}
We note that, due to the dependence of $e_{0}$ on
$v^{2}=\dot{x}^{2}$, \eqn{x-overd} is coupled to \eqn{e0}. Thus, the
overdamped \eqn{x-overd} could be also written in the form:
\begin{equation}
\label{x-overd-2}
\left(\gamma_{0} 
- d_2 \frac{q_{0}}{c+d_{2}\,\dot{x}^{2}} \right) \;\dot{x}
= F 
+ \sqrt{2k_{B}T\gamma_{0}} \; \xi(t)
\end{equation}
\eqn{x-overd-2} indicates a cubic equation for the velocities in the
overdamped limit, i.e. the possible existence of non-trivial solutions
for the stationary velocity. If we denote the stationary values of $v(t)$
by $v_{0}$ and neglect for the moment the stochastic term,
\eqn{x-overd-2} can then be rewritten as:
\begin{equation}
\Big[d_{2}\gamma_{0}\, v_0^2 -(q_{0} d_2
 -c\gamma_{0})\Big]\,  v_0 - d_2 F= c F. 
\label{v0-cubic}
\end{equation}
Depending on the value of $F$ and in particular on the sign of the term
\mbox{$(q_{0} d_2 -c\gamma_{0})$}, \eqn{v0-cubic} has either one or three
real solutions for the statio\-nary velocity, $v_{0}$. The always
existing solution expresses a direct response to the force in the form:
\begin{equation}
  \label{v-norm}
  v_{0} \sim F
\end{equation}
This solution results from the analytic continuation of Stokes' law,
$\bbox{v}_{0}= \bbox{F}/\gamma_{0}$, which is valid for $d_{2}=0$.
We will denote this solution as the ``normal'' mode of
motion, since the velocity $v$ has the same direction as the force
$F$ resulting from the external potential $U(x)$. 

As long as the supply of the energy depot is small, we will also name the
normal mode as the \emph{passive mode}, because the particle is simply
driven by the external force.  More interesting is the case of three
stationary velocities, $v_{0}$, which significantly depends on the
(supercritical) influence of the energy depot. In this case which will be
also discussed in the following sections, the particle will be able to
move in a ``high velocity'' or \emph{active mode} of motion.

For the one-dimensional motion, in the active mode only two different
directions are possible, i.e. a motion into or \emph{against} the
direction of the force $F$. 
Considering the linear potential $U(x)$, \eqn{u-x-f}, 
it is obvious that the particle's motion ``downhill'' is stable, but the
same does not necessarily apply for the possible solution of an
``uphill'' motion. Thus, in addition to \eqn{v0-cubic} which provides
the \emph{values} of the stationary solutions, we need a second condition
which guarantees the \emph{stability} of these solutions. 
In \cite{fs-tilch-eb-99}, we have investigated the necessary conditions
for such a motion in a linear potential. For the \emph{deterministic} case,
we found the following critical condition a possible ``uphill'' motion of
the pumped Brownian particles:
\begin{equation}
  \label{d2-f}
  d_{2}^{crit}=\frac{F^{4}}{8 q_{0}^{3}}\left(
1+\sqrt{1+\frac{4 \gamma_{0} q_{0}}{F^{2}}}\right)^{3}
\end{equation}
$d_{2}$ is the conversion rate of internal into kinetic energy, and
$F=-\nabla U$ describes the constant slope of the potential. In
the limit of negligible internal dissipation $c\to 0$, \eqn{d2-f}
describes how much power has to be supplied by the internal energy depot
to allow 
a \emph{stable uphill motion} of the particle.

In the following section, we will apply these results to a more
sophisticated potential. But before, we want to give a formal solution
for the Fokker-Planck \eqn{fpe-d2e} for the considered case $F=const.$ If
we neglect the space coordinate and concentrate on the  distribution
function $p(v,e,t)$ instead, the stationary solution $p^{0}(v,e)$ results
from the following equation: 
\begin{eqnarray}
\label{fpe-0f}
\frac{\partial}{\partial v} \left\{
\Big(\gamma_{0}-d_{2}e\Big) \,v\,
p(v,e) + D \, 
\frac{\partial p(v,e)}{\partial v} -
F\,p(v,e) \right\} = \non  
\frac{\partial}{\partial e} \Big[
q_{0} -c\, e -d_{2}v^{2}e\Big]\,p(v,e)
\end{eqnarray}
If we fix the depot energy $e$ to the value $e_{0}$, \eqn{e0}, the
stationary solution reads for sufficiently weak forces $F$: 

\begin{equation}
  \label{p0(v,e)}
  p^{0}(v,e) \sim \delta\Big(e-e_{0}\Big)\, \Big(c+d_{2}v^{2}
\Big)^{\frac{q_{0}}{2D}}\,\exp{\left[-\frac{\gamma_{0}}{2D}\,
\left(v-\frac{F}{\gamma_{0}}\right)^{2}\right]}
\end{equation}
where $\delta(e-e_{0})$ is Dirac's delta function. An investigation of
the possible stationary states in the full $\{v,e\}$ phase space is given
in \cite{fs-tilch-eb-99}

\section{Pumped Brownian Motion in a Ratchet Potential}
\label{3}
\subsection{Investigation of Deterministic Phase-Space Trajectories}
\label{3.1}
\begin{figure}[ht]
  \centerline{\psfig{figure=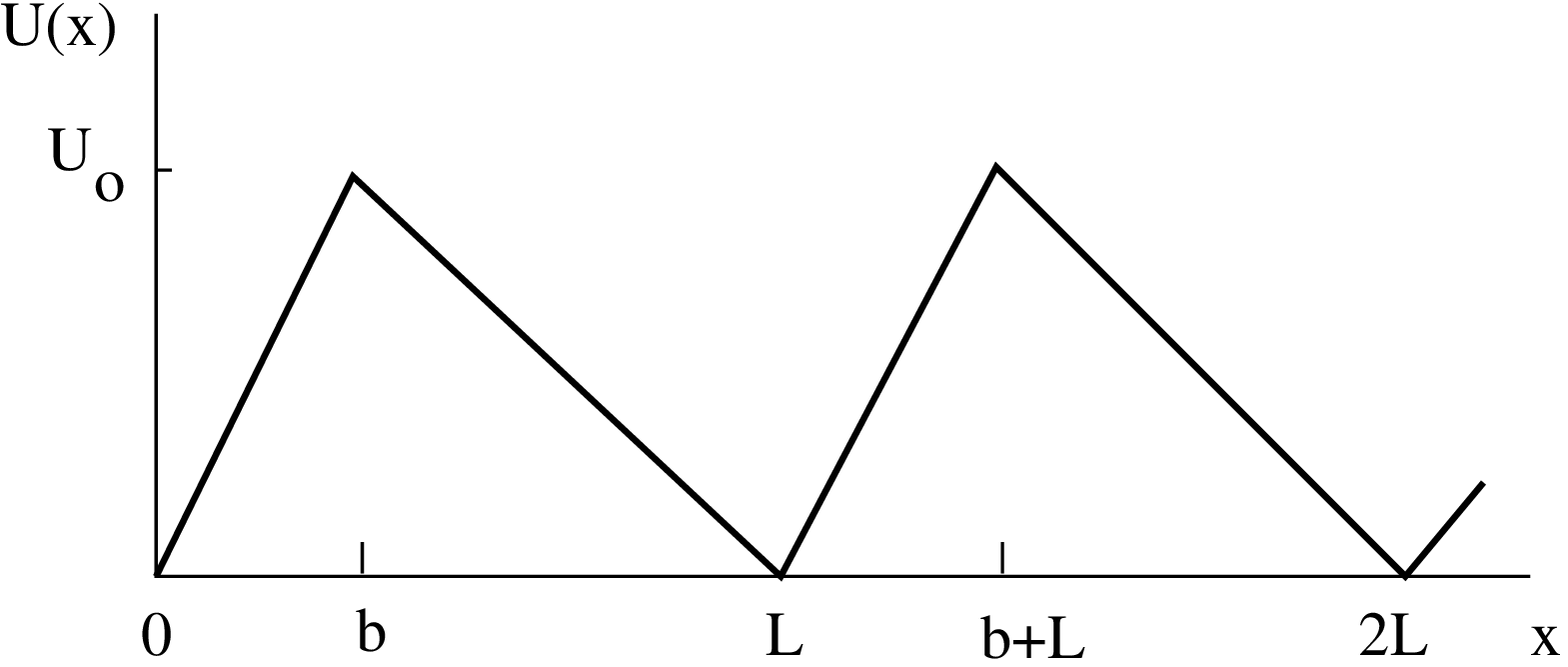,width=7.5cm,angle=-90}}
\caption[fig1]{
  Sketch of the asymmetric potential $U(x)$ (eq. (\ref{ux})). For the
  computer simulations, the following values are used: b=4, L=12, $U_{0}=
  7$ in arbitrary units.
\label{potent}}
\end{figure}

For further investigations of the motion of pumped particles by means of
computer simulations, we specify the potential $U(x)$ as a piecewise
linear, asymmetric potential (cf. Fig. \ref{potent}), which is known as a
\emph{ratchet potential}:
\begin{equation}
U(x)=\left\{
\begin{array}{lll}
\frac{\D U_0}{\D b}\{x-nL\} & \mbox{if $nL\leq x \leq nL+b$}
\\ & \mbox{$(n=0,1,2,...)$}\\
\frac{\D U_0}{\D L-b}\{(n+1)L-x\}\quad&\mbox{if $nL+b \leq x \leq (n+1)L$}
\\ & \mbox{$(n=0,1,2,...)$}\\
\end{array}
\right.
\label{ux}
\end{equation}
Further, we will use the following abbreviations with respect to the
potential $U(x)$, \eqn{ux}. The index $i=\{1,2\}$ refers to the two
pieces of the potential, $l_{1}=b$, $l_{2}=L-b$. The asymmetry parameter
$a$ should describe the ratio of the two pieces, and $F= -\nabla U =
const.$ is the force resulting from the gradient of the piecewise linear
potential. Hence, for the potential $U(x)$, \eqn{ux}, the following
relations yield:
\begin{eqnarray}
  \label{f-a}
 F_{1} &=& -\frac{U_{0}}{b}\;;\quad F_{2}=\frac{U_{0}}{L-b}\;;\quad 
a=\frac{l_{2}}{l_{1}}=\frac{L-b}{b}=-\frac{F_{1}}{F_{2}} \nonumber \\
F_{1}&=&-\frac{U_{0}}{L}(1+a)
\;;\quad F_{2} = \frac{U_{0}}{L}\frac{1+a}{a}
\end{eqnarray}
The motion of the Brownian particles with internal energy depot is still
described by the set of equations, (\ref{model}). In order to elucidate
the class of possible solutions for the dynamics specified, let us first
discuss the phase-space trajectories for the \emph{deterministic} motion,
i.e $D=0$. Further, $\mu=1$ and $\eta=1$ is assumed. 

Due to friction, a particle moving in the ratchet potential, will
eventually come to rest in one of the potential wells, because the
dissipation is not compensated by the energy provided from the internal
energy depot. The series of \pic{phase-traj} shows the corresponding
attractor structures for the particle's motion dependent on the supply of
energy expressed in terms of the conversion rate $d_{2}$.  In
\pic{phase-traj}a, we see that for a subcritical supply of energy
expressed in terms of the conversion rate $d_{2}$ only \emph{localized}
states for the particles exist. The formation of limit cycles inside each
minimum corresponds to stable oscillations in the potential well, i.e.
the particles are not able to escape from the potential well.

With increasing $d_{2}$, the particles are able to climb up the potential
flank with the lower slope, and this way escape from the potential well
into negative direction. As \pic{phase-traj}b shows,
this holds also for particles which initially start into the positive
direction. Thus, we find an unbound attractor corresponding to
\emph{delocalized motion} for negative values of $v$. Only if the
conversion rate $d_{2}$ is large enough to allow the uphill motion along
the flank with the steeper slope, the particles can escape from the
potential well in \emph{both} directions, and we find two unbound
attractors corresponding to \emph{delocalized} motion
into  both positive and negative direction. 

To conclude these investigations, we find that the structure of the phase
space of a active Brownian particle in an unsymmetrical ratchet potential
may be rather complex. While at low pumping rates bound attractors (limit
cycles) in each potential well are observed, with increasing pumping
rates one/two new unbound attractors are formed, which correspond to a
directed stationary transport of the particles into negative/positive
direction.

Considering the threedimensional phase space, $\{x,v,e\}$, 
these two stationary solutions are separated by a two-dimensional
separatrix plane. This plane has to be periodic in space because of the
periodicity of the ratchet potential. In order to get an idea of the
shape of the separatrix we have performed computer simulations which
determined the direction of motion of one particle for various initial
conditions. %
\pic{separat} shows the respective trajectories for the movement in both
directions and the separatrix plane.  We can conclude that, if a particle
moves into the positive direction, most of the time the trajectory is
very close to the separatrix. That means it will be rather susceptible
for small perturbations, i.e.  even small fluctuations might be able to
destabilize the motion into the positive direction. The motion into
negative direction, on the other hand, is not susceptible in the same
manner, since the respective trajectory remains in a considerable
distance from the separatrix or comes close to the separatrix only for a
very short time. This conclusion will be of importance when
discussing the current reversal in a ratchet potential in \sect{4}. In
particular, in \sect{4.1} we will investigate the existence of critical
conversion rates $d_{2}$ in more detail by comparing analytical results
and computer simulations.

\subsection{Deterministic vs. Stochastic Motion}
\label{3.2}

In the following, the motion of an {\em ensemble} of $N$ pumped Brownian
particles in a ratchet potential, \eqn{ux}, is investigated both for the
deterministic and the stochastic case. For the computer simulations, we
have assumed that the start locations of the particles are equally
distributed over the first period of the potential, $\{0,L\}$. We note,
that in the computer simulations always the complete set of equations
(\ref{model}) for the particles is solved.

In the \emph{deterministic case}, the direction of motion and the
velocity at any time $t$ are mainly determined by the initial conditions.
Provided a sufficient supply of energy, particles with an initial
position between $\{0,b\}$, which initially feel a force into the
negative direction, most likely move with a negative velocity, whereas
particles with an initial position between $\{b,L\}$ most likely move
into the positive direction. For parameter values which allow the
existence of two unbound attractors (cf. also \pic{phase-traj}c),
\pic{damp-ve-x0} shows the final distribution of the velocities in the
deterministic case.  
We see two main currents of particles occuring, with a positive and a
negative velocity. 
The net current, however, has a positive direction, since most of the
particles start with the matching initial condition.  The time dependence
of the averages is shown in the first part of \pic{damp-noi-avest} for
$t\leq 2.000$. The long-term oscillations in the average velocity and the
average energy depot result from the superposition of the velocities,
which are sharply peaked around the two dominating values, shown in
\pic{damp-ve-x0}.

In order to demonstrate the influence of fluctuations on the mean values
$\la x \ra$, $\la v \ra$, $\la e \ra$, we add a stochastic force with
$D>0$ to the simulation of the ensemble of pumped Brownian particles
shown in \pic{damp-noi-avest} at $t=2.000$. Hence, the second part of
\pic{damp-noi-avest} demonstrates the changes after the stochastic force
is switched on. The simulations show that the current changes its
direction when noise is present. Different from the deterministic case,
\pic{damp-ve-x0}, the related velocity distribution in the stochastic
case now approaches an one peak distribution, shown in
\pic{damp-noi-ve-vert}. 
Hence, stochastic effects are able to stabilize the motion in the
vicinity of the unbound attractor which corresponds to the negative
current, while they destabilize the motion in the vicinity of the unbound
attractor corresponding to the positive current, both shown in
\pic{phase-traj}c.

A similar situation can be also observed when the motion of the particles
is less damped (cf. \pic{und-ensem-xve-t}). For this case, we clearly see
both the mean velocity and the mean energy depot approaching constant
values, instead of oscillating. Further, the time scale of the
relaxation into the stationary values decreases with the intensity of the
stochastic force.

The related velocity distribution of the particles is plotted for
different intensities of the stochastic force in \pic{und-ve-vert}. For
$D=0$, there are two sharp peaks at positive values for the speed and one
peak for negative values, hence, the average current is positive. The
three maxima correspond to the stationary velocities in the deterministic
case. For $D=0.001$, the distribution of positive velocities is
declining, since most of the particles already move with a negative
velocity, and for $D=0.01$ after the same time all particles have a
negative velocity. Further, the velocity distribution becomes more
broaden in the presence of noise.

As a result of the computer simulations presented in this section, we
find that an ensemble of pumped Brownian particles moving in a ratchet
potential is able to produce a directed net current. In the deterministic
case, we find two currents into opposite directions related to a sharply
peaked bimodal velocity distribution, and the direction of the resulting
net current is determined by the initial conditions of the majority of
the particles. In the stochastic case, however, we find only a broad and
symmetric unimodal velocity distribution, resulting in a stronger net
current. The direction of this net current is opposite to the
deterministic case and points into the negative direction, hence,
stochastic forces are able to change the direction of motion of the
particles moving into the positive $x$-direction. This reversal will be
discussed in more detail in the next section.

\section{Investigation of the Current Reversal}  
\label{4}
\subsection{Dependence on Energy Conversion and Noise}
\label{4.1}

The previous computer simulations revealed the dependence of the net
current on the initial conditions and on the influence of the stochastic
force. Since all particles have started in the first period of the
ratchet potential, the existence of periodic stationary solutions,
$v_{0}(x) = v_{0}(x \pm L)$, requires that the particles are able to
escape from the potential well.  Hence, they must be able to move
``uphill'' on one or both flanks of the ratchet potential, in order to
obtain a net current. In the phase space, this corresponds to the
formation of unbound attractors corresponding to directed motion in the
ratchet potential. This, in turn, would require a supercritical supply of
energy for the pumped Brownian particles to move in the active or ``high
velocity'' mode, as already mentioned in \sect{2}.

\eqn{d2-f} provides a necessary condition for a stable uphill motion on a
single flank in terms of a critical conversion rate $d_{2}^{crit}$. 
In order to demonstrate the applicability of \eqn{d2-f} for the ratchet
potential, we have investigated the dependence of the \emph{net current},
expressed by the mean velocity $\mean{v}$, on the conversion rate,
$d_{2}$. The deterministic motion is for the overdamped case discussed in
\cite{fs-tilch-eb-99}, here we concentrate on the less damped case and
discuss both the determinstic and the stochastic motion.  The results of
computer simulations are shown in \pic{v-d2-stoch}.

For the \emph{determinstic motion}, we see the existence of \emph{two
  different critical values} for the parameter $d_{2}$, which correspond
to the onset of a \emph{negative net current} at $d_{2}^{crit1}$ and a
\emph{positive net current} at $d_{2}^{crit2}$.  For values of $d_2$ near
zero and less than $d_{2}^{crit1}$, there is no net current at all. This
is due to the subcritical supply of energy from the internal depot, which
does not allow an uphill motion on any flank of the potential.
Consequently, after the initial downhill motion, the particles perform a
localized motion (oscillation) within the potential well (cf. also
\pic{phase-traj}a) and no current occurs. The amplitude of these
oszillations depends on the conversion rate $d_2$. In the limit of
vanishing energy conversion the motion of the particles comes to rest in
the minima of the ratchet potential.

With an increasing value of $d_{2}$, we see the occurence of a negative
net current at $d_{2}^{crit1}$. That means, the energy depot provides
enough energy for the uphill motion along the flank with the lower slope,
which, in our example, is the one with $F=7/8$ (cf.  \pic{potent}). If we
insert this value for $F$ into the critical condition, \eqn{d2-f}, a
value $d_{2}^{crit1}=1.03$ is obtained, which agrees with the onset of
the negative current in the deterministic computer simulations,
\pic{v-d2-stoch}.

For $d_{2}^{crit1} \leq d_{2}\leq d_{2}^{crit2}$, a stable motion of the
particles up and down the flank with the lower slope is possible, but the
same does not necessarily apply for the steeper slope. Hence, particles
which start on the lower slope with a positive velocity, cannot continue
their motion into the positive direction since they are not able to climb
up the steeper slope. Consequently, they turn their direction on the
steeper slope, then move downhill driven by the force into the negative
direction, and continue to move into the negative direction while
climbing up the lower slope. Therefore, for values of the conversion rate
between $d_{2}^{crit1}$ and $d_{2}^{crit2}$, we only have an
\emph{unimodal} distribution of the velocity, and only one unbound
attractor exists as shown in \pic{phase-traj}b.

For $d_{2}>d_{2}^{crit2}$, the energy depot also supplies enough energy
for the particles to climb up the steeper slope, consequently a periodic
motion of the particles into the positive direction becomes possible,
now. In our example, the steeper slope corresponds to the force $F=-7/4$
(cf. \pic{potent}) which yields a critical value $d_{2}^{crit1}=11.3$,
obtained by means of \eqn{d2-f}. This result agrees with the onset of the
positive current in the computer simulations, \pic{v-d2-stoch}.  For
$d_{2}>d_{2}^{crit2}$, we have a \emph{bimodal} velocity distribution in
the deterministic case, as also shown in the top part of
\pic{und-ve-vert}. This corresponds to the existence of two unbound
attractors as shown in \pic{phase-traj}c. For equally distributed initial
positions of the particles, the net current which results from the
average of the two main currents has a positive direction in the
deterministic case, because most of the particles start into a positive
direction, as discussed above.

Let us now turn to the \emph{stochastic case}. In \sect{3}, we have
already shown that the influence of a stochastic force may lead to a
\emph{current reversal}. In the deterministic case, the particles will
keep their direction determined by the initial conditions provided the
energy supply allows them to move ``uphill'', which is the case for
$d_{2}>d_{2}^{crit2}$. In the stochastic case, however, the initial
conditions will be ``forgotten'' after a short time, hence due to
stochastic influences, the particle's ``upill'' motion along the steeper
flank will soon turn into a ``downhill'' motion. This motion into the
negative direction will be most likely kept because less energy is
needed. Thus, the stochastic fluctuations reveal the instability of an
``uphill'' motion along the steeper slope.

For a comparison with the determinsitic case, \pic{v-d2-stoch} shows the
average velocity also for the stochastic case.  Here, the net current is
always negative in agreement with the explanation above. This holds even
if the supercritical supply of energy, expressed by the conversion
parameter, $d_{2}>d_{2}^{crit2}$, would allow a \emph{deterministic}
motion into the positive direction (cf. the dashed line in
\pic{v-d2-stoch}). In addition, we find a \emph{very small} positive net
current in the range of small $d_{2}$ (cf.  the insert in
\pic{v-d2-stoch}). Whereas in the deterministic case, for the same values
of $d_{2}$ no net current at all is obtained, the fluctuations in the
stochastic case allow some particles to escape the potential barriers.

In order to investigate how much the strength $D$ of the stochastic force
may influence the magnitude of the net current into the negative
direction, we have varied $D$ for a fixed conversion parameter
$d_{2}=1.0$ for the less damped case. As \pic{v-d2-stoch} indicates, for
this setup there will be only a negligible net current, $\mean{v}\approx
0$ in the deterministic case ($D=0$), but a remarkable net current,
$\mean{v}=-0.43$ in the stochastic case for $D=0.1$. As \pic{v-s} shows,
there is a \emph{critical strength} of the stochastic force,
$D^{crit}(d_{2}=1.0)\simeq 10^{-4}$, where an onset of the net current
can be observed, while for $D<D^{crit}$ no net current occurs. On the
other hand, there is also an \emph{optimal strength} of the stochastic
force, $D^{opt}$, where the amount of the net current, $\abs{\mean{v}}$,
reaches a \emph{maximum}. An increase of the stochastic force above
$D^{opt}$ will only increase the \emph{randomness} of the particle's
motion, hence the net current decreases again. In conclusion, this
sensitive dependence on the stochastic force may be used to adjust a
\emph{maximum net current} for the particles movement in the ratchet
potential.

\subsection{Dependence on Potential Asymmetry}
\label{4.2}

We conclude our results by investigating the influence of the slope on
the establishment of a positive or negative net current. With a fixed
height of the potential barrier $U_{0}$, and a fixed length $L$, the
ratio of the two different slopes is described by the asymmetry parameter
$a$ of the ratchet potential, \eqn{f-a}.
 In the deterministic case, the occurence of a current in
the ratchet potential 
depends on the critical supply of energy, described by \eqn{d2-f}.  In
order to obtain the critical value for the asymmetry of the potential, we
replace the force $F$ in \eqn{d2-f} by the parameter $a$, \eqn{f-a}. In
our example, the flank  $l_{1}$ of the potential has the steeper slope,
so the critical condition is determined by $F_{1}=U_{0}/L\;(1+a)$.  As
the result, we have found \cite{fs-tilch-eb-99}:
\begin{equation}
  \label{a-crit1}
  a^{crit}=\frac{L}{U_{0}}\left[-\frac{\gamma_{0}}{2} d_{2}^{-1/3} +
\sqrt{\frac{\gamma^{2}}{4} d_{2}^{-2/3}+q_{0}d_{2}^{1/3}}
\;\right]^{3/2} -1 
\end{equation}
$a^{crit} \geq 1$ gives the critical value for the asymmetry, which may
result in a reversal of the net current in the deterministic case. 

\pic{v-a-stoch} shows the average velocity, $\mean{v}$, dependent on the
asymmetry parameter, $a$.  The simulations have been carried out for
different values of the stochastic force in the \emph{overdamped case},
for a fixed value $d_{2}=10$. In particular, the top part of
\pic{v-a-stoch} also includes the curve for $D=0$. As we see in the
deterministic case, for $a>a^{crit}$ the flank $l_{1}$ is too steep for
the particles, therefore only a negative current can occur.
For $1<a<a^{crit}$, however, the particles are able to move uphill either
flank. Hence, also a positive net current can establish.

As long as the stochastic forces are below a critical value,
$D<D^{crit}$, which is about $D^{crit}(d_{2}=10)\simeq 0.02$, the curves
for the stochastic case are not very different from the deterministic
one, as shown in the top part of \pic{v-a-stoch}. Hence, the above
conclusions about the current reversal in the deterministic case apply.
However, for $D>D^{crit}$ (cf. the bottom part of \pic{v-a-stoch}) we do
not find a critical asymmetry $a^{crit}$ for a current reversal.
Instead, the net current keeps its positive (for $a<1$) or negative (for
$a>1$) direction for any value of $a<1$ or $a>1$, respectively.

The bottom part of \pic{v-a-stoch} further indicates the existence of an
optimal strength of the stochastic force, again. Similar to the
investigations in \pic{v-s}, we find that an increase in $D$ does not
necessarily results in a increase of the amount of the net current. In
fact, the maximum value of $\abs{\mean{v}}$ is smaller for $D=0.1$ than
for $D=0.05$, which indicates an optimal strength of the stochastic
force, $D^{opt}$, between $0.05$ and $0.1$ for the considered set of
parameters.

We note that the values for both the critical strength, $D^{crit}$, and
the optimal strength, $D^{opt}$, of the stochastic force depend on the
value of the conversion parameter, $d_{2}$, as a comparison of \pic{v-s}
and \pic{v-a-stoch} shows. Further, these values may be also functions of
the other parameters, such as $q_{0}$, $\gamma_{0}$, and therefore differ
for the strongly overdamped and the less damped case.

\section{Conclusions}
\label{5}

In Sects. \ref{3} and \ref{4}, we have shown by means of computer
simulations, that an ensemble of pumped Brownian particles moving in a
ratchet potential can produce a \emph{directed net current}.
Hence, by means of an appropriate \emph{asymmetric potential} and an
additional mechanism to drive the system into \emph{non-equilibrium}, we
are able to convert the genuinely \emph{non-directed} Brownian motion
into \emph{directed motion}.  In this respect, our result agrees with the
conclusions of other physical ratchet models which have been proposed to
reveal the microscopic mechanisms resulting in directed movement. In
order to point ot some differences to our model, we first need to
summarize the basic principles.

Ratchet models which describe the directed transport of particles, are
usually based on \emph{three ingredients}: (i) an \emph{asymmetric
  periodic potential}, known as ratchet potential, \eqn{ux},
\pic{potent}, which lacks the reflection symmetry, (ii) \emph{stochastic
  forces} $\xi(t)$, i.e. the influence of noise resulting from
thermal fluctuations on the microscale, and (iii) \emph{additional
  correlations}, which push the system \emph{out of thermodynamic
  equilibrium}. For the latter one, different assumptions can be made.
In the example of the {\em flashing ratchet}, the correlations result
from an fluctuating energy profile. In the overdamped limit, the dynamics
of a Brownian particle can then be described by the equation of
motion:
\begin{equation}
  \label{flash}
\frac{dx}{dt} = - \zeta(t) \left. \frac{\partial U(x)}{\partial x}
\right.
+ \sqrt{2 D}\;\xi(t)
\end{equation}
The \emph{nonequilibrium forcing} $\zeta(t)$ which governs the time
dependent change of the potential, can be either considered as a
periodic, deterministic modulation with period $\tau$, $\zeta(t) \to
F(t)=F(t+\tau)$, or as a stochastic \emph{non-white} process $\zeta(t)$. In
the special case where $\zeta(t)$ or $F(t)$ have only the values
$\{0,1\}$, the periodic potential is switched $ON$ and $OFF$. 

In another class of ratchet models, the additional correlations result
from \emph{spatially uniform forces} of temporal or statistical
\emph{zero average}. In the example of the \emph{rocking ratchet}, the particles are subject to a spatially uniform
time-periodic deterministic force $F(t)=F(t+\tau)$, for instance:
\begin{equation}
  \label{rock}
\frac{dx}{dt} = - \left. \frac{\partial U(x)}{\partial x} \right.
- A\,\cos{\left(\frac{2\pi\,t}{\tau}\right)} + \sqrt{2 D}\;\xi(t)
\end{equation}
Here, the potential $U_{s}(x,t)=U(x)+A\,x\,\cos{(\Omega\,t)}$ is
periodically rocked. If
the value of $A$ is adjusted properly, in the deterministic case a net
current of particles into the direction of the \emph{lower slope} can be
observed, while the movement into the direction of the steeper slope is
still blocked.  However, a consideration of stochastic influences results
in a \emph{current reversal} for a certain range of the parameters $A$
and $D$ \cite{Bartussek-Haenggi-Kissner-94}. Then, the net current occurs again in the direction of
the \emph{steeper slope}. Because this phenomenon depends also on the
mass of the particles, it can be used for mass separation
\cite{Lindner-et-97}. 

Another example of the same class, the {\em correlation ratchet}
\cite{magnasco-93,luczka-bart-haenggi-95,millonas-dykman-94}, is also
driven by a spatially uniform, but stochastic force, $\zeta(t)$. Here,
the equation for the overdamped motion reads for instance:
\begin{equation}
  \label{correl}
\frac{dx}{dt} = - \left. \frac{\partial U(x)}{\partial x} \right.
+ \zeta(t) + \sqrt{2 D}\;\xi(t)
\end{equation}
where $\zeta(t)$ is a time correlated (colored) noise of zero average. As
a third example, the {\em diffusion ratchet} \cite{Reimann-et-96} is
driven by a spatially uniform, time-periodic diffusion coefficient
\mbox{$D(t)=D(t+\tau)$}, which may result e.g. from an oscillating
temperature.  In the overdamped limit, the equation of motion reads for
example:
\begin{equation}
  \label{diffus}
\frac{dx}{dt} = - \left. \frac{\partial U(x)}{\partial x} \right.
+ \zeta(t) + \Big[1+\,A\,sin(\Omega\,t)\Big]\; \sqrt{2 D}\;\xi(t)
\end{equation}
Inspite of the different ways to introduce additional correlations, all
these models have in common to transfer the undirected motion of Brownian
particles into directed motion, hence, the term {\em Brownian rectifiers}
\cite{haenggi-bart-96} has been established.

As we have shown in this paper, our own model can also serve for this
purpose. For a comparison with the models above, we rewrite \eqn{x-overd}
for the overdamped motion of the Brownian particles with internal energy
depot in a more general way:
\begin{equation}
  \label{x-sumar}
\frac{dx}{dt} = - f(t)\;\left. \frac{\partial U(x)}{\partial x} 
\right.
+ C\,f(t)\; \sqrt{2D} \;\xi_i(t)\;\;; \quad
f(t) =  \frac{1}{\gamma_{0}-d_{2}e(t)}
\end{equation}
In \eqn{x-sumar}, the prefactor $f(t)$ appears
\emph{twice}, up to a constant $C$: it changes \emph{both} the influence
of the ratchet potential, and the magnitude of the diffusion coefficient.
In \emph{this} respect, our model is between a flashing ratchet model,
\eqn{flash}, and a diffusion ratchet model, \eqn{diffus}.

$f(t)$ is \emph{in general} a time dependent
function, because of $e(t)$. In the limit of a stationary
  approximation, $e(t)\to e_{0}$, \eqn{e0}, we found
\cite{fs-tilch-eb-99} that $f(t)$ may switch between two constant
  values $f_{1}(x)>0$, $f_{2}(x)<0$:
\begin{equation}
\label{pref-1}
f_{i}(x) 
= \frac{1}{2\gamma_{0} F_{i}} \left(F_{i}
\pm \sqrt{F_{i}^2+4q_0\gamma_{0}}\right) 
\end{equation}
dependent on the moving direction and the flank, the particle is moving
on. Hence, the function $f(t)$ does not represent a spatially uniform
force. However, it has been shown in \cite{fs-tilch-eb-99}, that for the
stationary approximation $\mean{\gamma_{0} -d_2e_{0}} \,\tau = 0$ holds,
i.e. the force is of zero average with respect to one period, $\tau$.
Hence, we conclude that the mechanism of motion which \eqn{x-overd} is
based on, should be different from the previous mechanisms which
originate directed motion in a ratchet potential.

Finally, we note that the model of active Brownian particles with
internal energy depot is not restricted to ratchet systems. In a general
sense, these particles can be described as Brownian machines or {\em
  molecular motors} which convert chemical energy into mechanical motion.
While different ideas for Brownian machines have been suggested
\cite{magnasco-94,astumian-bier-94,juelicher-prost-95}, our model aims to
add a new perspective to this problem. The ability of the particles to
take up energy from the environment, to store it in an internal depot and
to convert internal energy to perform different activities, may open the
door to a more refined description of microbiological processes based on
physical principles.




\section*{Figures}

\vfill

\begin{figure}[ht]
  \centerline{\psfig{figure=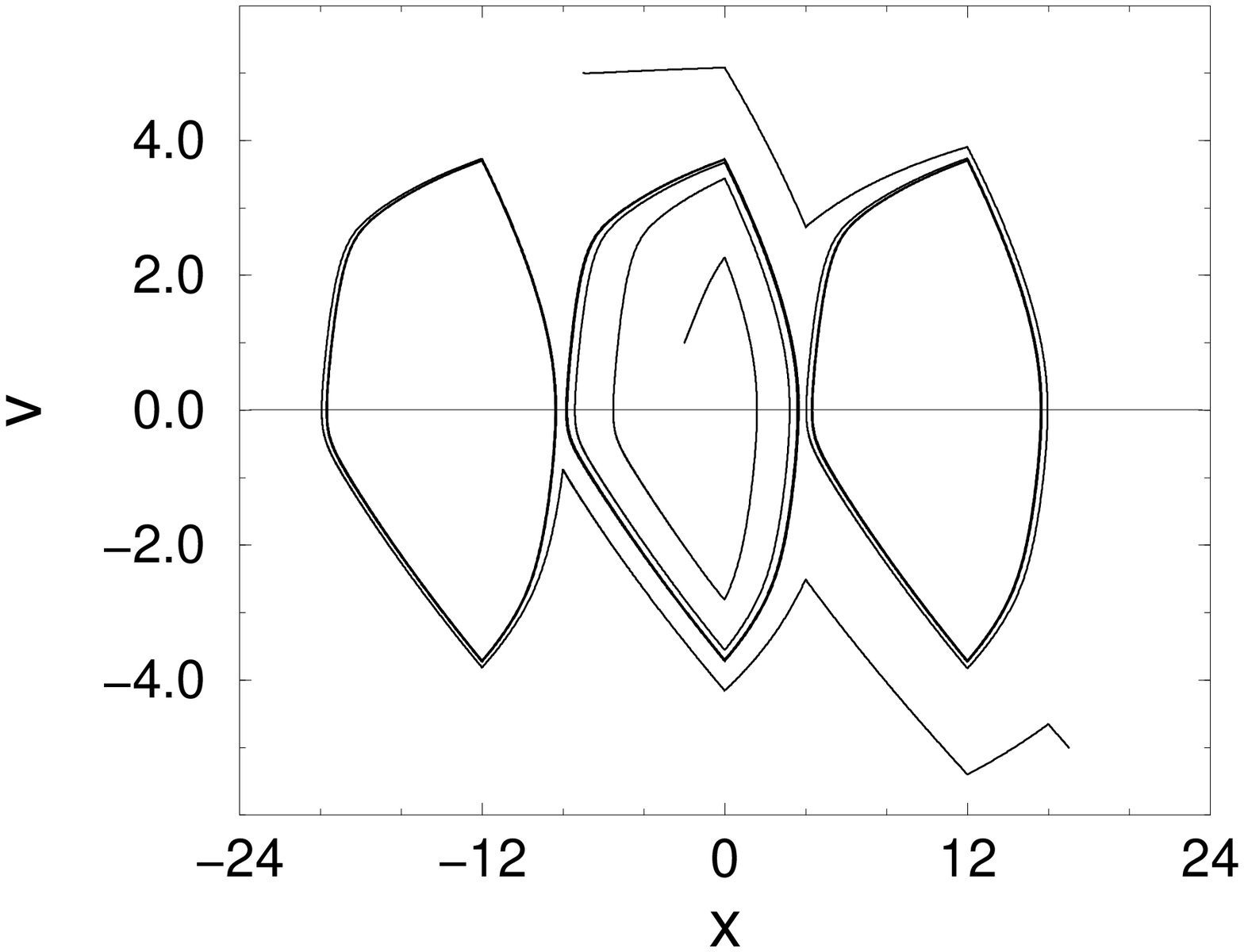,width=7.5cm}}

  \centerline{\psfig{figure=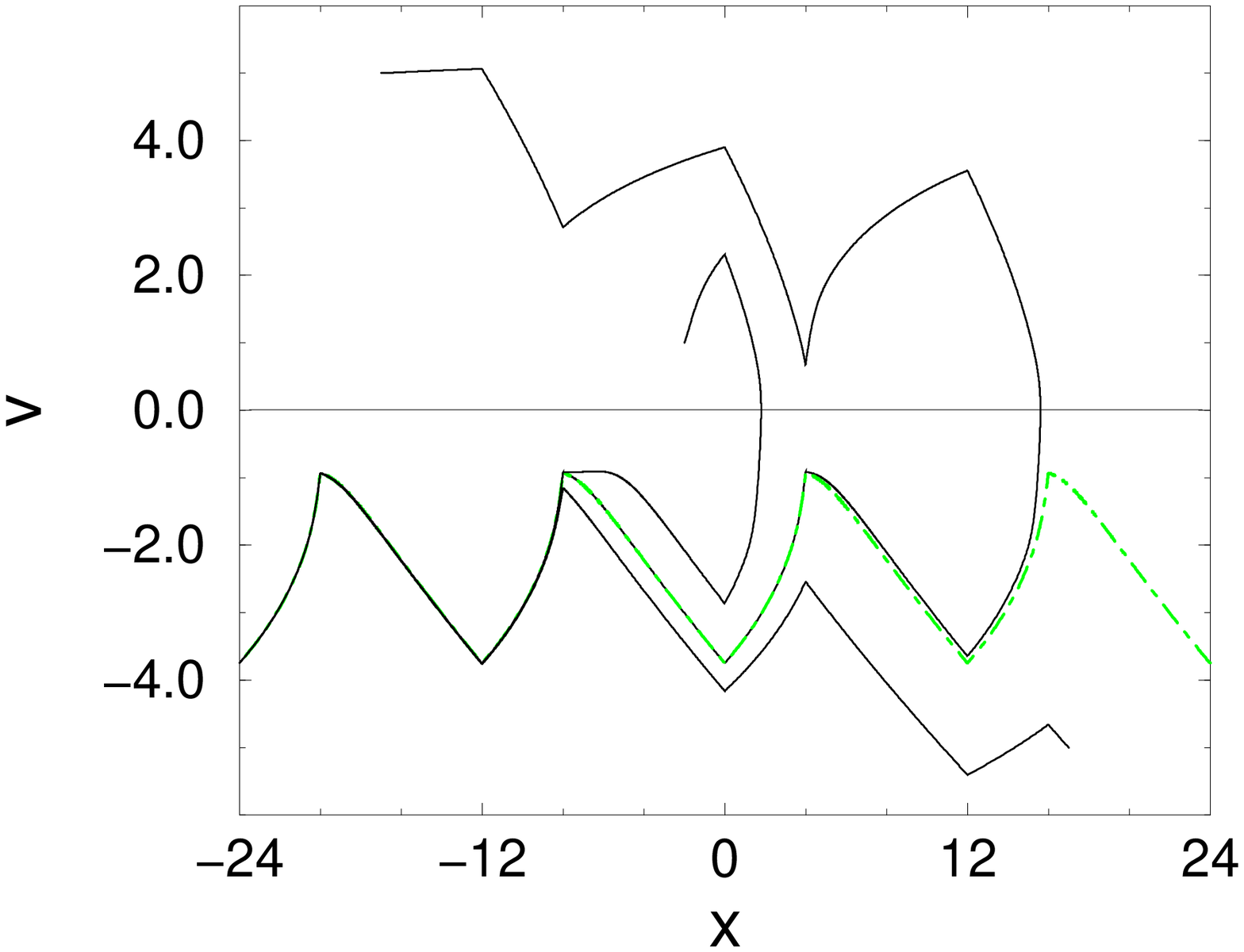,width=7.5cm}}

  \centerline{\psfig{figure=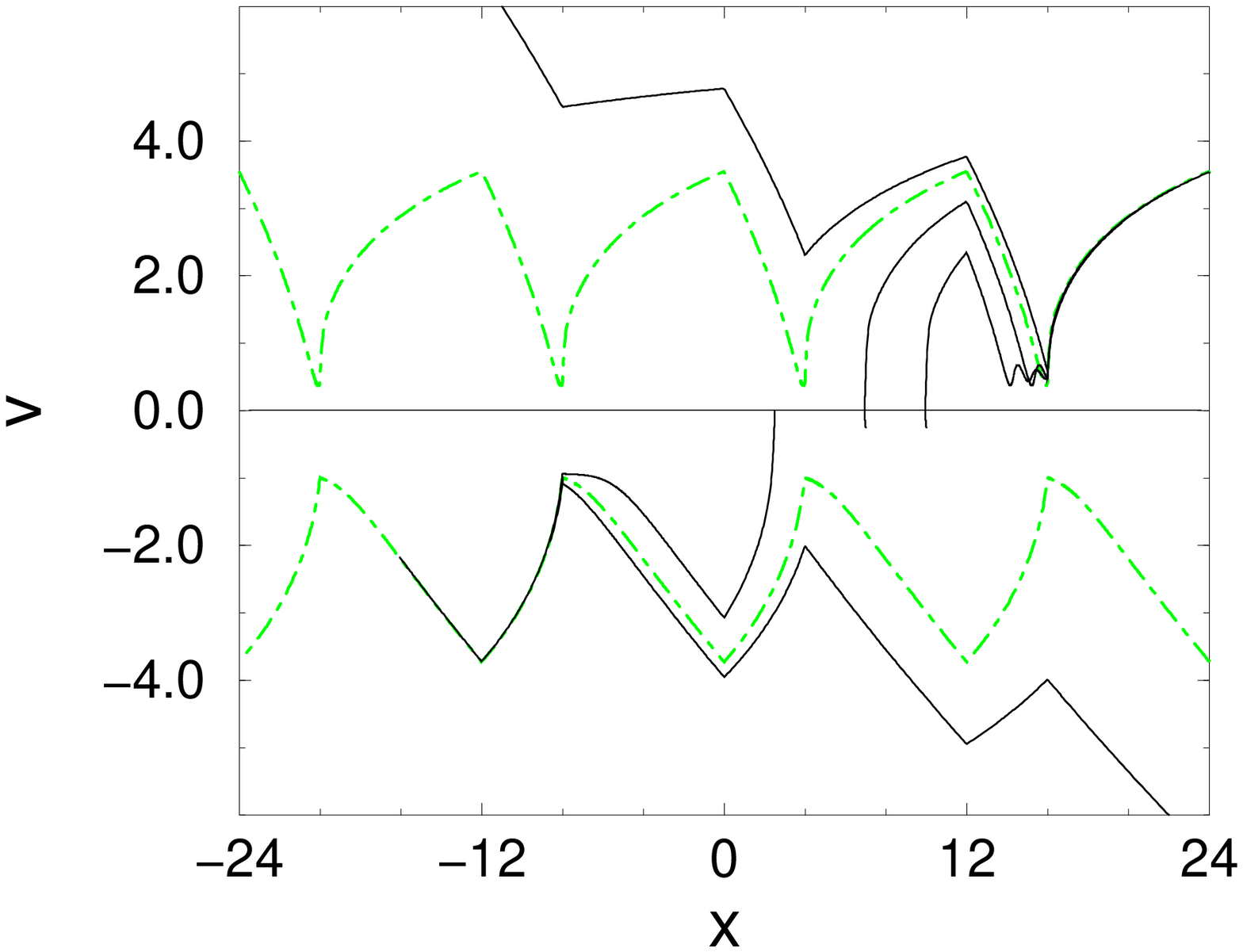,width=7.5cm}}
\caption[fig1]{
  Phase-space trajectories of particles starting with different intitial
  conditions, for three different values of the conversion parameter
  $d_{2}$: (a: top) $d_{2}=1$, (b: middle) $d_{2}=4$, (c: bottom)
  $d_{2}=14$.  Other parameters: $q_{0}=1$, $c=0.1$, $\gamma_{0}=0.2$.
  The dashed-dotted lines in the middle and bottom part show the unbound
  attractor of the delocalized motion which is obtained in the long-time
  limit.
\label{phase-traj}}
\end{figure}

\begin{figure}[ht]
\centerline{\psfig{figure=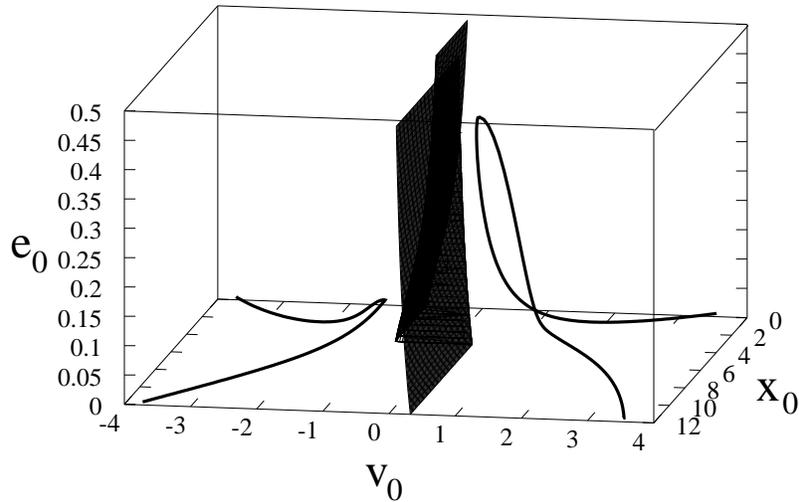,width=10.5cm}}
\caption[fig4]{
Separatrix and asymptotic trajectories in the $\{x,v,e\}$ phase
space. Parameters see \pic{phase-traj}c.  In order to obtain the
separatrix in \pic{separat}, we have computed the velocity $v_0$ for
given values of $x_0$ and $e_0$ (in steps of 0.01) using the Newton
iteration method with a accuracy of 0.0001.
\label{separat}}
\end{figure}

\begin{figure}[ht]
\centerline{\psfig{figure=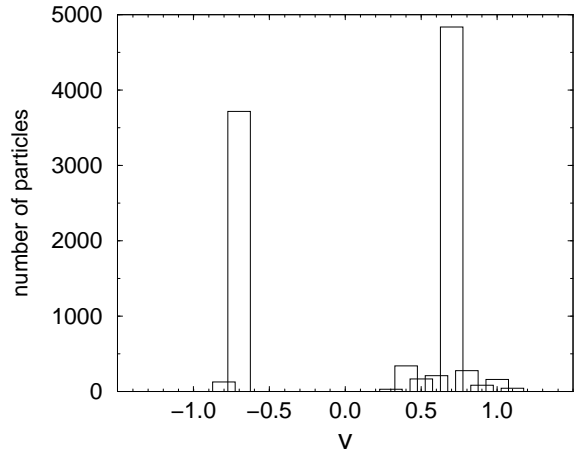,width=7.5cm}}
\caption[fig4]{
  Distribution of the final velocity $v_{e}$ after $t=2.000$ simulation
  steps (averaged for 10.000 particles). The initial locations of the
  particles are equally distributed over the first period of the ratchet
  potential $\{0,L\}$). Parameters: $q_0=10, \gamma=20, c=0.01, d_2=10$
  (strongly damped case).  \label{damp-ve-x0}}
\end{figure}

\vfill

\begin{figure}
\centerline{
  \psfig{figure=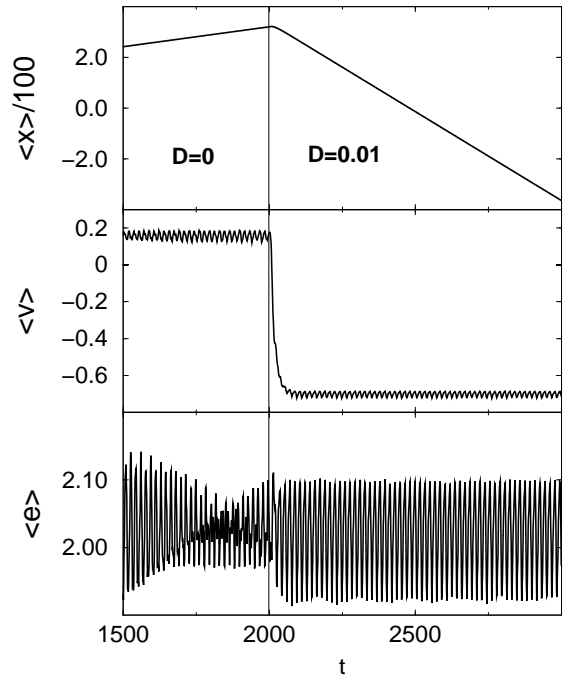,width=7.5cm}}
  \caption[fig6]{
    Averaged location $\langle x\rangle$, velocity $\langle v \rangle$
    and energy depot $\langle e \rangle$ versus time $t$ for 10.000
    particles. The stochastic force $D=0.01$ is switched on at $t=2.000$.
    Parameters see \pic{damp-ve-x0}. \label{damp-noi-avest}}
\end{figure}

\vfill

\begin{figure}
\centerline{\psfig{figure=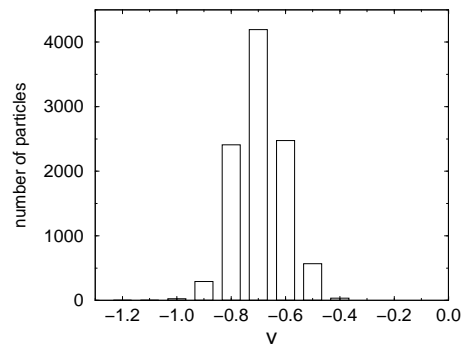,width=6.cm}}
\caption[]{
  Distribution of velocity $v$ for the simulation of \pic{damp-noi-avest}
  at time $t=4.000$, which means 2.000 simulation steps
  after the stochastic force was switched on.
  \label{damp-noi-ve-vert}}
\end{figure}

\vfill

\begin{figure}[ht]
\centerline{
  \psfig{figure=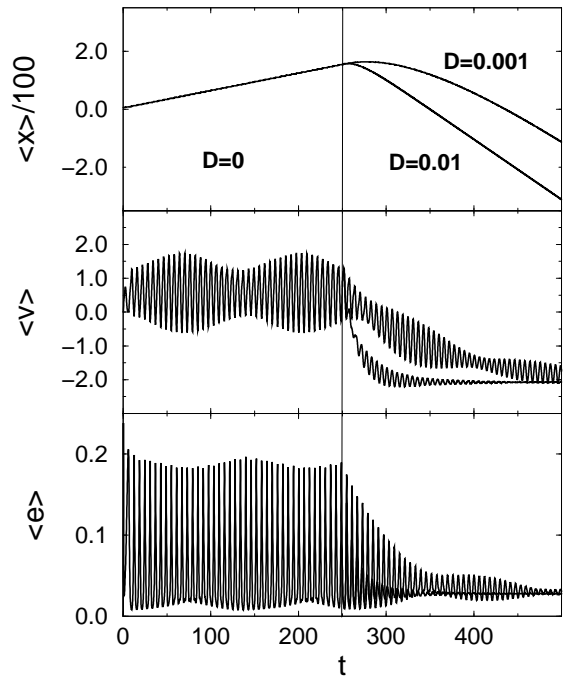,width=7.5cm}}
  \caption[fig8]{
    Averaged location $\langle x\rangle$, velocity $\langle v \rangle$
    and energy depot $\langle e \rangle$ versus time $t$ for 10.000
    particles.  The stochastic force, either $D=0.01$ or $D=0.001$, is
    switched on at $t=250$.  Parameters: $q_0=1.0, \gamma = 0.2, c=0.1,
    d_2=14.0$ (less damped case). \label{und-ensem-xve-t}}
\end{figure}

\clearpage 

\begin{figure}[ht]
\centerline{\psfig{figure=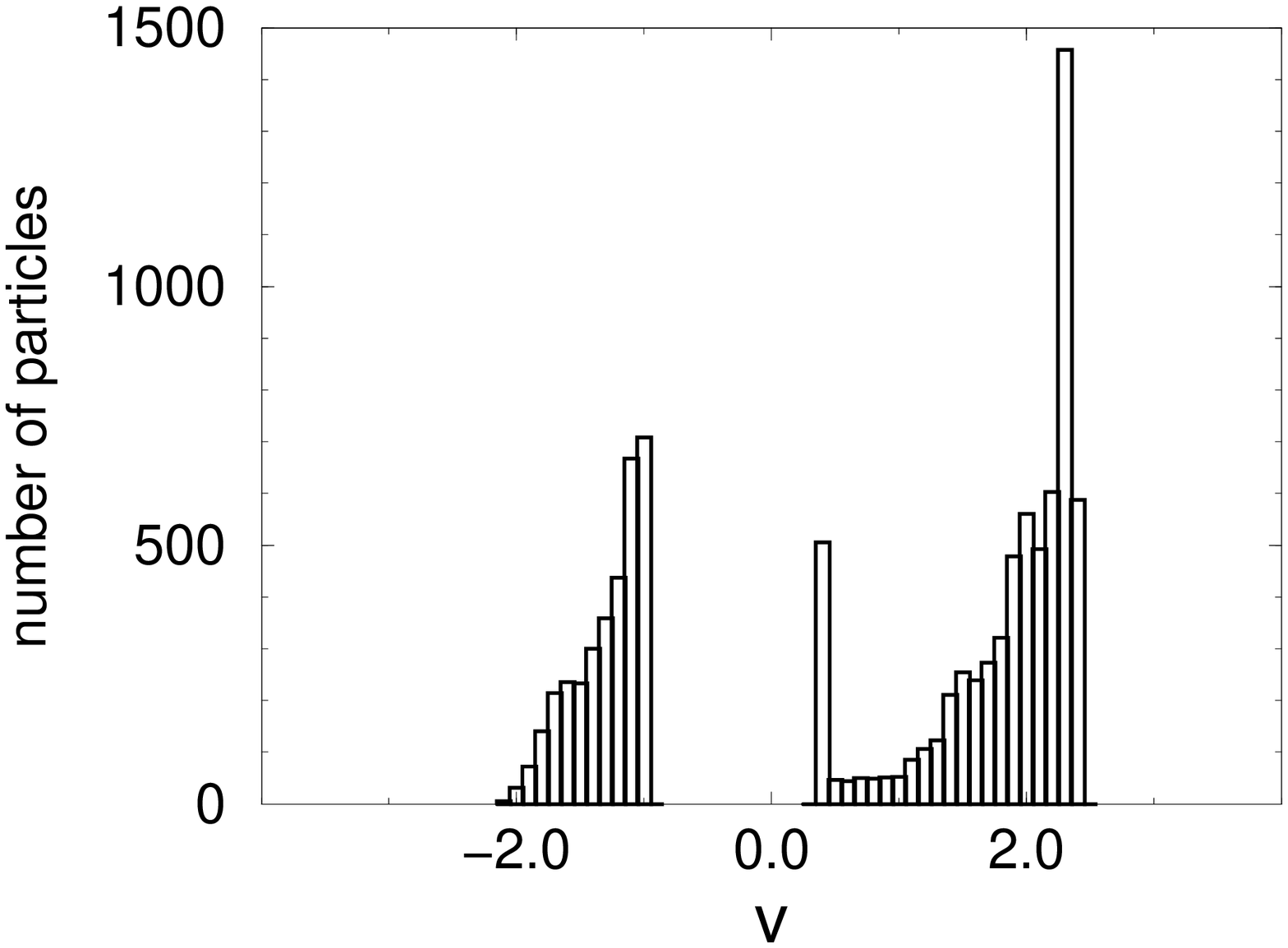,width=7.5cm}}
\centerline{\psfig{figure=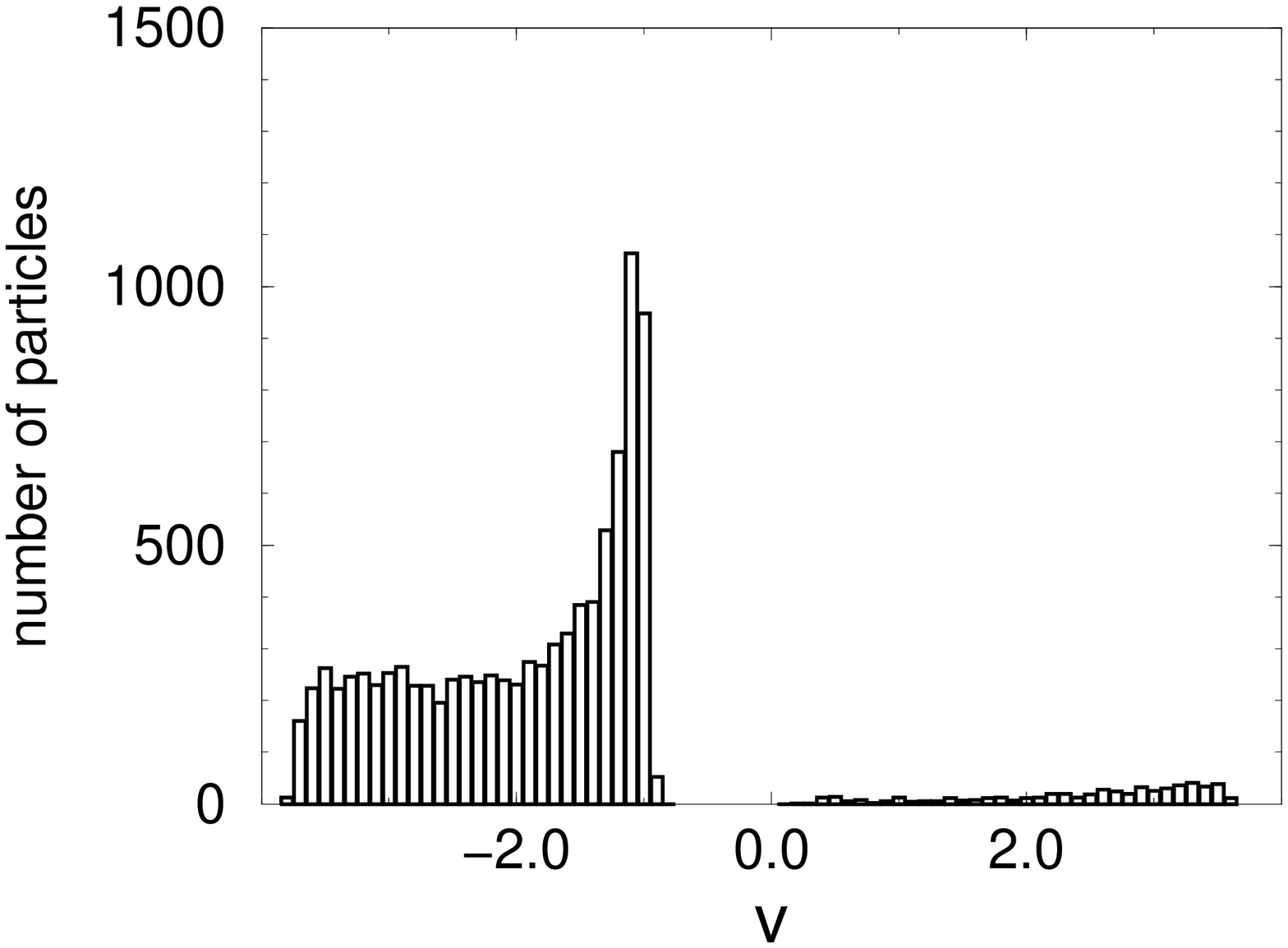,width=7.5cm}}
\centerline{\psfig{figure=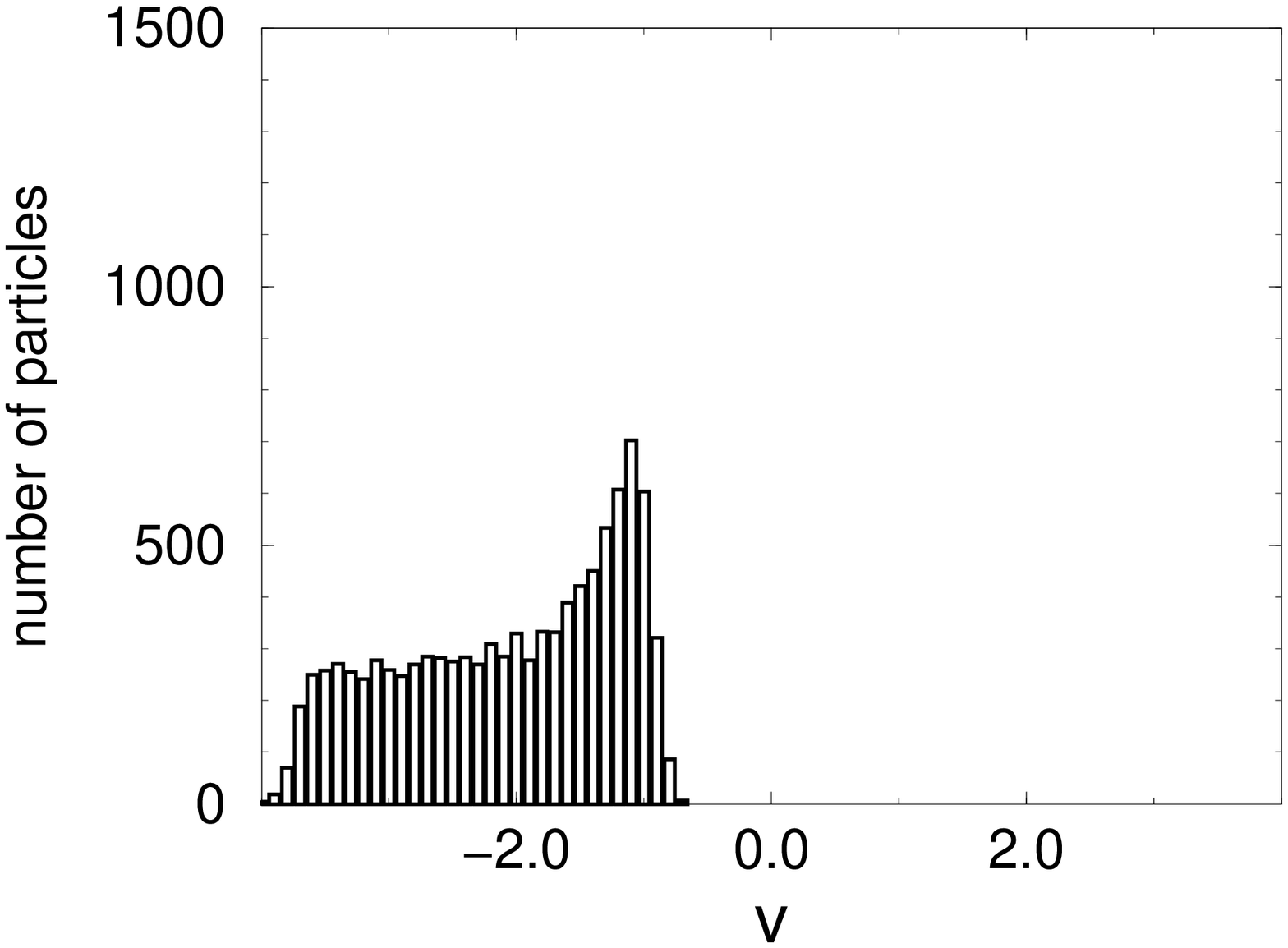,width=7.5cm}}
\caption[fig9]{
  Distribution of velocity $v$ for the simulation of
  \pic{und-ensem-xve-t}. (top) $D=0.0$ ($t=250$),
  (middle) $D=0.001$ ($t=500$), (bottom) $D=0.01$ ($t=500$). The
  stochastic force is switched on at $t = 250$.
    \label{und-ve-vert}}
\end{figure}

\vfill

\begin{figure}[htbp]
  \centerline{\psfig{figure=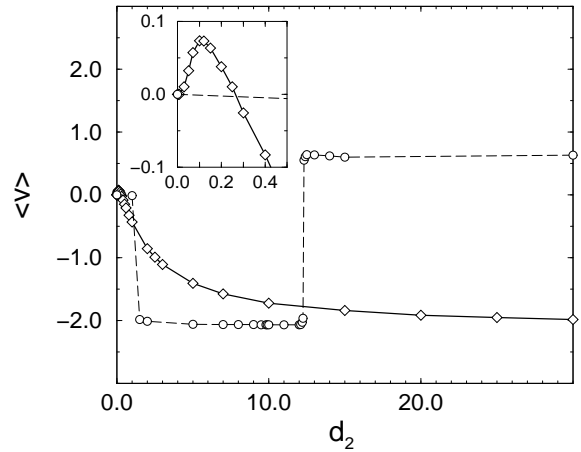,width=7.5cm}}
\caption{
  Average velocity $\langle v \rangle$ vs. conversion parameter $d_2$.
  The data points are obtained from simulations of 10.000 particles with
  arbitrary initial positions in the first period of the ratchet
  potential. $(\Diamond)$ stochastic case ($D=0.1$), $(\circ)$
  deterministic case ($D=0$), for the other parameters see
  \pic{und-ensem-xve-t}.  \label{v-d2-stoch}}
\end{figure}

\vfill

\begin{figure}[htbp]
  \centerline{\psfig{figure=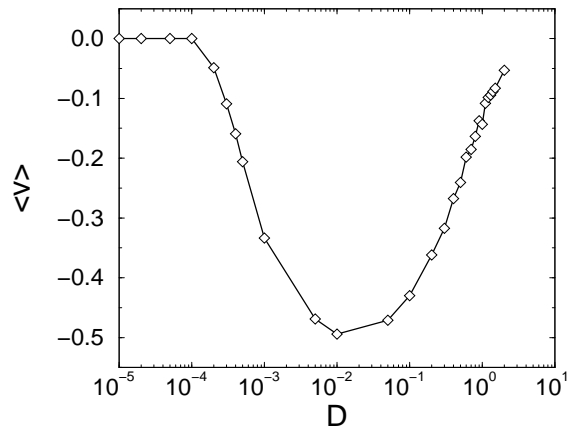,width=7.5cm}}
\caption{
  Average velocity $\langle v \rangle$ vs. strength of the stochastic
  force $D$.  The data points are obtained from simulations of 10.000
  particles with a fixed conversion parameter $d_2=1.0$, for the other
  parameters see \pic{und-ensem-xve-t}. \label{v-s}}
\end{figure}

\vfill

\begin{figure}[htbp]
 \centerline{\psfig{figure=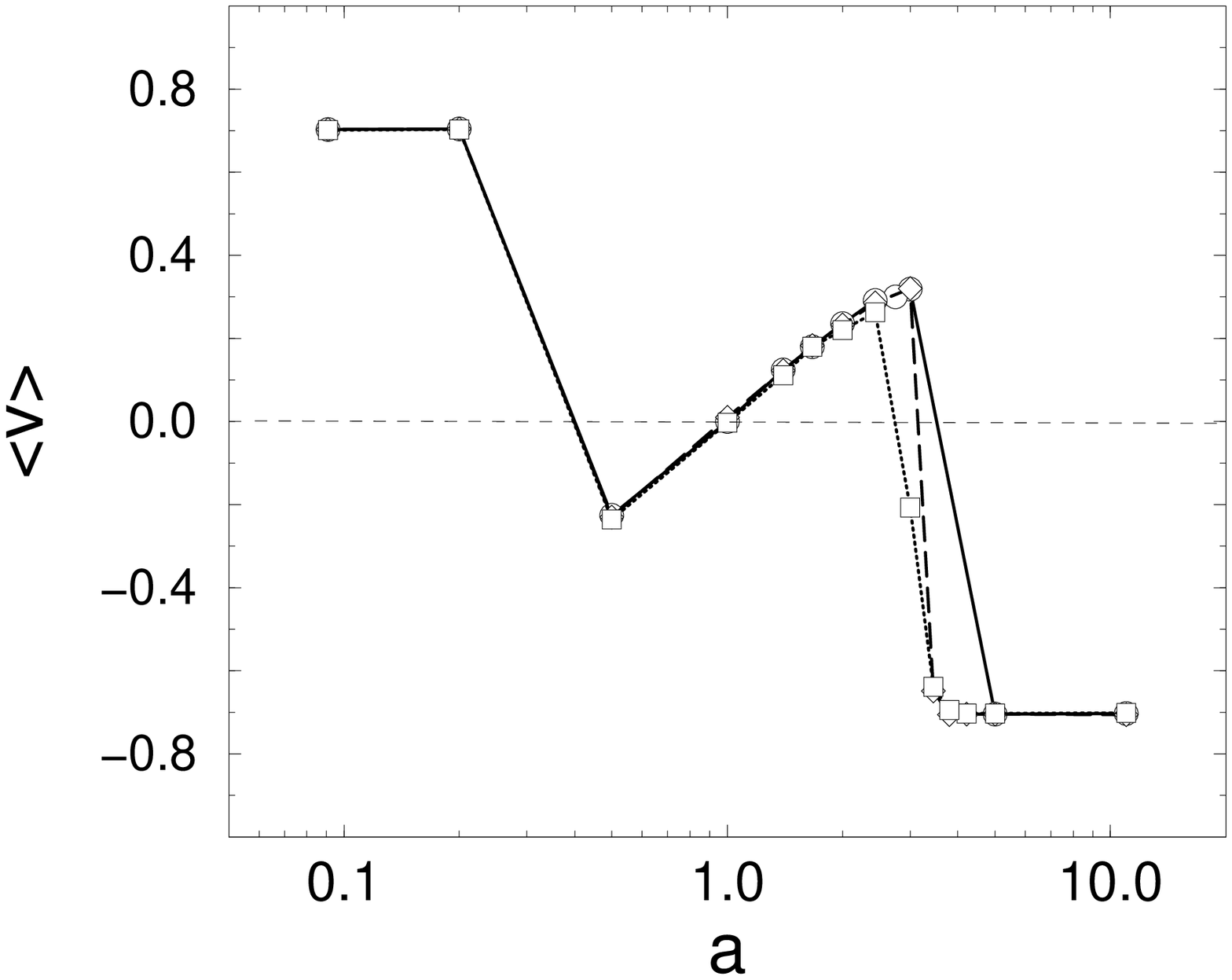,width=7.5cm}} 

 \centerline{\psfig{figure=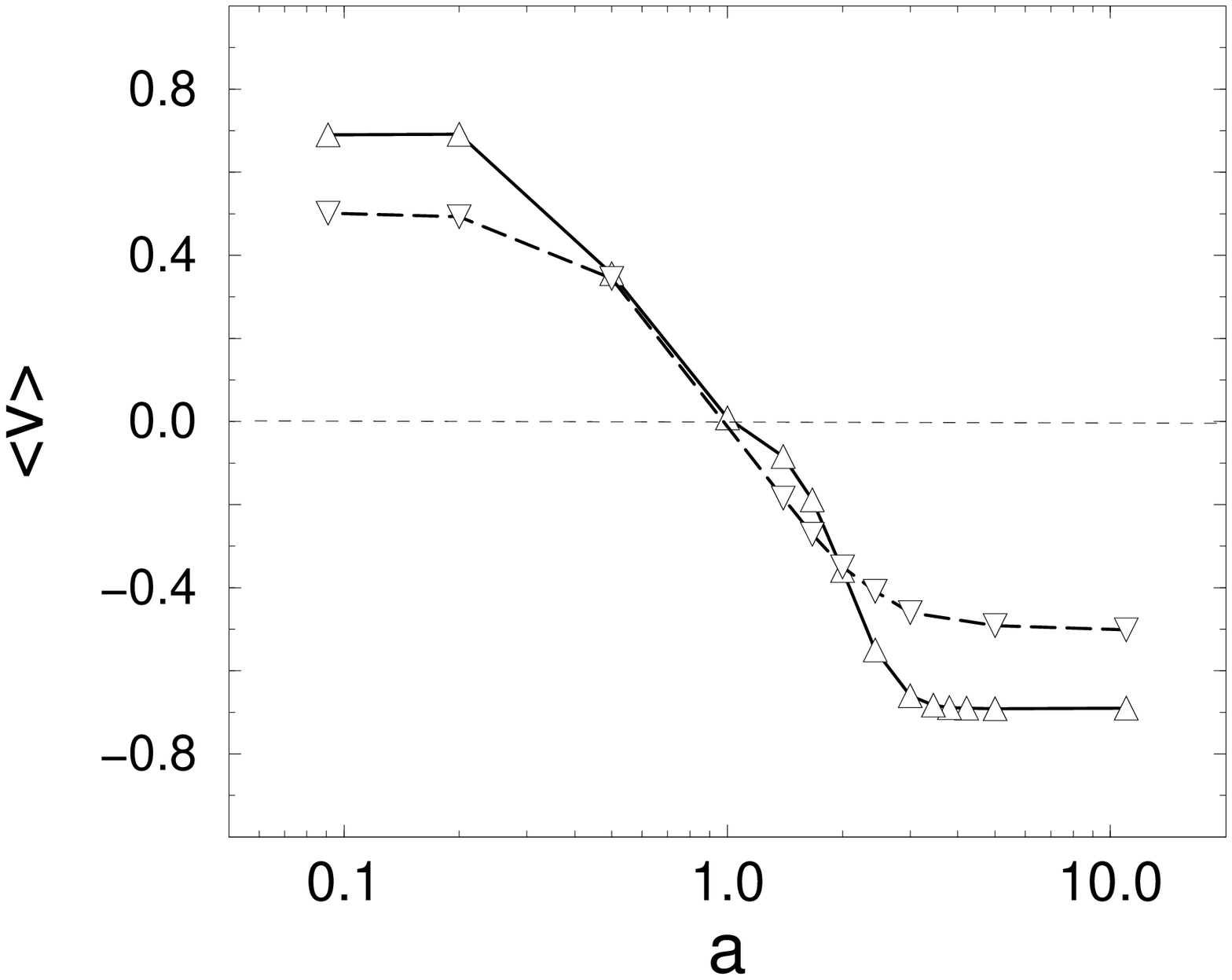,width=7.5cm}}
\caption{
  Average velocity $\mean{v}$ vs. asymmetry parameter $a$, \eqn{f-a}.
  The data points are obtained from simulations of 10.000 particles with
  arbitrary initial positions in the first period of the ratchet
  potential. (top) subcritical stochastic force: $(\circ)$ $D=0$
  $(-\!\!\!-\!\!\!-)$, $(\Diamond)$ $D=0.001$ $(--)$, $(\Box)$ $D=0.01$
  $(\cdots)$, (bottom) supercritical stochastic force: $(\bigtriangleup)$
  $D=0.05$ $(-\!\!\!-\!\!\!-)$, $(\bigtriangledown)$ $D=0.1$ $(--)$.
  $d_{2}=10$, for the other parameters see \pic{damp-ve-x0}.
  \label{v-a-stoch}}
\end{figure}

\end{document}